# Phase transition close to room temperature in BiFeO$_3$ thin films

J. Kreisel[*], P. Jadhav, O. Chaix-Pluchery

1   Laboratoire des Matériaux et du Génie Physique, Grenoble INP, CNRS, Minatec, 3, parvis Louis Néel, 38016 Grenoble, France

M. Varela

2   Dept Física Aplicada i Òptica, Universitat de Barcelona, Carrer Martí i Franquès 1. 08028 Campus UAB, Bellaterra 08193, Spain

N. Dix, F. Sánchez and J. Fontcuberta

3   Institut de Ciència de Materials de Barcelona (ICMAB–CSIC), Campus UAB, Bellaterra 08193, Spain

**Abstract**

BiFeO$_3$ (BFO) multiferroic oxide has a complex phase diagram that can be mapped by appropriately substrate-induced strain in epitaxial films. By using Raman spectroscopy, we conclusively show that films of the so-called supertetragonal T-BFO phase, stabilized under compressive strain, displays a reversible temperature-induced phase transition at about 100 ºC, thus close to room temperature.

* Corresponding author:   *jens.kreisel@grenoble-inp.fr*

The room-temperature antiferromagnetic (AFM) and ferroelectric (FE) BiFeO$_3$ (BFO) perovskite oxide is currently the workhorse of most advanced research in multiferroic materials, which attract a great deal of interest [1-3]. Its rich and complex phase diagrams are at the root of its capability to be used in novel devices either for electric control of magnetic materials, for lead-free ferroelectrics or novel piezoelectric materials with giant response [4]. Whereas room-temperature bulk BFO has a *G*-type AFM rhombohedral *R*3*c* structure, it has been shown that strained BFO films can display distinct symmetries and polarizations. In particular, it has been shown that compressive stress reduces the symmetry by producing a remarkable elongation of the out-of-plane cell parameter, thus an increased *c/a* ratio (pseudocubic notation is used throughout this paper) and giving rise to the so-called supertetragonal T-BFO phase with high polarization [5, 6]. Although the symmetry of strained films is neither purely rhombohedral or tetragonal but more likely monoclinic [7-9], it is customary to label the corresponding phases as *R*-like and *T*-like to emphasize the symmetry of the corresponding parent structures. Zeches *et al.*[10] have reported that both *R*-like and *T*-like phases may coexist in strained films, similarly to the morphotropic phase boundary (MPB) between *R*-like and *T*-like phases in other composition-modulated perovskite oxides with a large piezoelectric response [11, 12].

The occurrence of these coexisting phases is conditioned by the fact that they are almost degenerated under strain [8]. In addition, *ab-initio* calculations have shown that there are a number of distinct monoclinic *T*-like phases (i.e. *Pc*, *Cm*, *Cc*) that compete each other and thus the actual stabilized phase can critically depend on strain, growth conditions and temperature [8]. Whereas the relative stability under strain has been addressed by exploring BFO films coherently grown on mismatched substrates [10, 13-17], the temperature dependence of compressively strained T-BFO films remains less explored.

Here we will report on the growth of BFO films on compressing (001) oriented LaAlO$_3$ (LAO) substrates (-4.3% mismatch respect to rhombohedral BFO) and their characterization, with a focus on the investigation by Raman spectroscopy (RS) of the thermal stability. RS is known to be a versatile probe for the investigation of structural and physical properties in BFO single crystals, ceramics, thin films and nanostructured samples [9, 18-24]. We will show that the room-temperature stabilized *T*-like phase displays a reversible structural phase transition at relatively low temperature (about 100 ºC) towards a new phase. This finding illustrates the richness of the BFO phase diagram and shows that new phases with potentially interesting properties can be obtained at the MPB within the T-like region, with new or enhanced properties.

BiFeO$_3$ films were deposited by pulsed laser deposition (KrF excimer laser) on LaAlO$_3$(001) substrates (a$_{pc}$ = 3.79Å) at 700 ºC and an oxygen background pressure of 0.1 mbar by ablating a target of Bi$_{1.1}$FeO$_3$ composition.

In **Fig 1(a)** we show the θ-2θ scan X-ray diffraction pattern of a 100 nm thick BFO film. We observe a series of well-defined and high-intensity peaks corresponding to (*00l*) reflections whose position (labeled T$_{00l}$) match those of the T-BFO phase, sometimes also known as M$_c$-phase [6, 8]. The corresponding *c*-axis cell parameter is $c_{T-BFO}$ ~ 4.655(8) Å. Similarly to reported data [13, 25], additional weak reflections are observed and are attributed to a minor fractions of monoclinic *R*-like BFO (labeled *M$_R$*) and relaxed bulk-like R-BFO (labeled *R*) domains. A more detailed study (**Fig. 1(b-d)**) using reciprocal space maps (rms) and rocking curves, confirms the prevalence of the T-BFO peak (relaxed R-like BFO spots are barely observed) with minor traces of M$_R$, tilted M$_R$ (M$_{Rt}$) and tilted T-BFO (T$_t$). The rocking curve around the T-BFO (002) reflection (**Fig. 1(c)**) reveals the presence of two shoulders caused by tilted T-like domains (with Δω ≈ ± 1.1-1.2º) [25, 26]. A splitting (Δω = 0.15º) in the T$_{002}$-reflection can also be observed; it likely originates from twinning induced by the LAO substrate. The shape of the {103} T-BFO spots in the rsm (Fig. 1(d)) is compatible with monoclinic M$_C$ structure reported in ref. [10, 13, 25, 26] using high-resolution diffractometers.

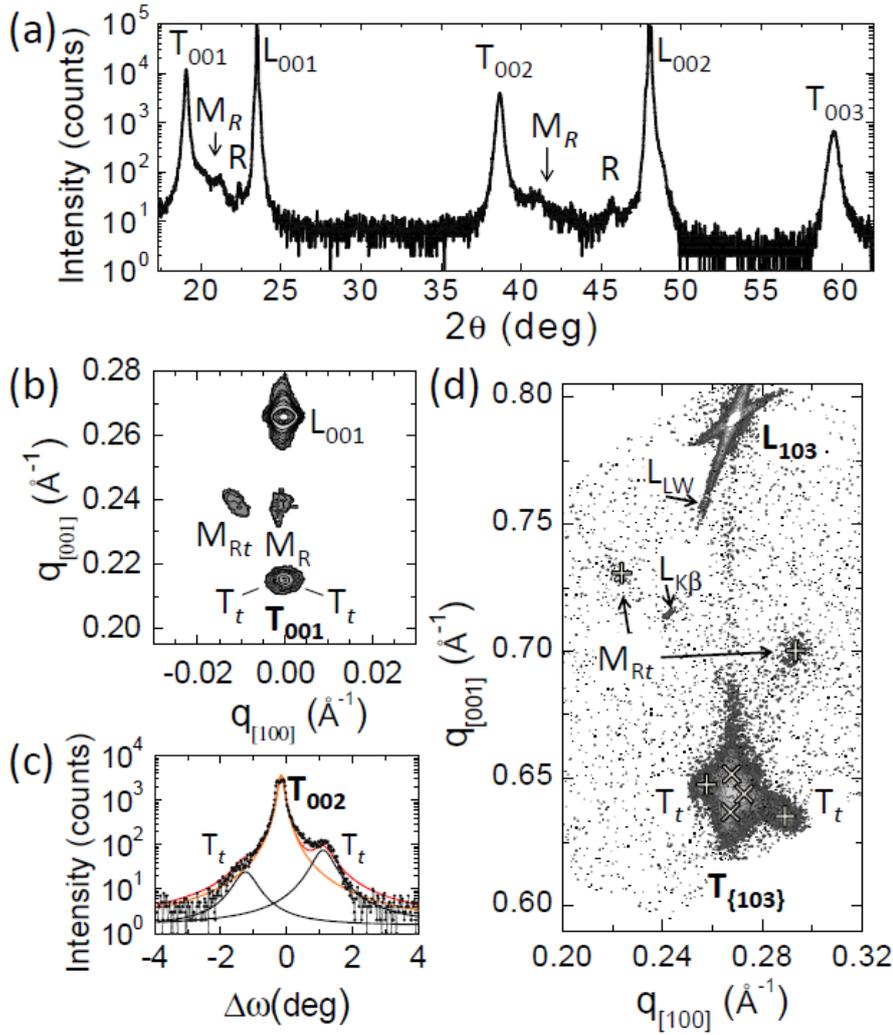

*Fig. 1 (colour on-line)*
**(a)** XRD θ/2θ scan of a 100 nm thick BFO film. (*00l*) peaks of T-like BFO, monoclinic *R*-like BFO and relaxed R-BFO are labeled $T_{00l}$, $M_R$ and R, respectively. $L_{(00l)}$ are substrate reflections. **(b)** Reciprocal space maps around the (001) substrate reflection showing the presence of tilted $M_R$ ($M_{Rt}$) domains. **(c)** ω-scan for the $T_{002}$-BFO reflection; the two shoulders at Δω ~ 1.1° indicate tilted *T*-like domains ($T_t$); **(d)** Reciprocal space maps around the (103) LAO substrate reflection (labeled $L_{(103)}$); peaks caused by spurious $Cu_{Kβ}$ and $W_L$ wavelengths are labeled $L_{Kβ}$, $L_{LW}$, respectively. The positions of the {013} reflections of tetragonal-like T-BFO, marked with $T_{\{103\}}$, are indicated by (x) symbols. The approximate peak positions of minor fraction of tilted $T_t$ and $M_{Rt}$ domains are also indicated (+).

The spots corresponding to the T-like phase ($M_c$) have been used to evaluate the lattice parameters: $a ≈ 3.75(1)$ Å, $b ≈ 3.80(1)$ Å, $c ≈ 4.66(1)$ Å and $β ≈ 88.6(3)°$. On the other hand, the out-of-plane lattice parameter of the minority phases $M_R$ and *R*-BFO phases $M_R$ and *R*-BFO phases, determined from data in Fig. 1(a) are: $c_{MR-BFO} = 4.19(2)$Å, $c_{R-BFO}=3.97(1)$Å, which are in good agreement with previous reports [25, 26]. In summary the XRD data described above confirms that our BFO//LAO film consists largely (abundance about 99%) of supertetragonal *T*-like T-BFO with c/a ≈ 1.23.

The film has then been investigated by Raman spectroscopy using a 514 nm laser line through a LabRam Jobin-Yvon spectrometer with a spectral cut-off at around 100 cm$^{-1}$. Spectra have been taken by using a laser power of less than 1 mW into a focused spot of about 2 µm$^2$ under the microscope to avoid sample heating. Controlled temperature-dependent Raman measurements from 23 °C to 600 °C have been carried out by using a commercial LINKAM THS600 heating stage placed under the Raman microscope. The Raman spectra before and after temperature measurements are identical, attesting the reversibility of temperature-induced changes.

**Fig. 2** presents selected temperature-dependent Raman spectra of the T-BFO film on LAO. At ambient conditions, the well-defined Raman spectrum is characterized by the superposition of sharp bands from the LAO substrate (120, 150 and 484 cm$^{-1}$) and bands originating from the BFO thin film. The BFO spectrum is characterized by four sharp and intense bands at 225, 266, 359 and 685 cm$^{-1}$, a further intense but large feature at 589 cm$^{-1}$ and several low-intensity features which are more difficult to assign. This signature is consistent with earlier reported 300 K spectra on T-BFO films [9, 27]. The distinctive characteristic of the T-BFO Raman spectrum is the prominent and sharp band at 685 cm$^{-1}$ which is not present in rhombohedral BFO films. Based on the number and symmetry of the observed modes, Iliev *et al* .[9, 27] have interpreted the Raman spectrum of T-BFO as a signature of a monoclinic *Cc* symmetry with a large *c/a*, ratio rather than a *P4mm* symmetry, in agreement with theoretical predictions [7, 8]. However, it has been noted that the Raman signature of T-BFO alone does not allow differentiating between the *Cc* and the alternatively suggested *Cm* structure [6], leaving the place for both hypotheses.

It is important to note that the room temperature Raman spectrum of our T-BFO thin film is entirely explained by the T-BFO signature without detectable traces of other phases, which is at variance with the reported spectral phase coexistence observed for thicker T-BFO films where the proportion of relaxed R-phase increases with thickness [9]. The observation of a pure T-BFO signature is in agreement the XRD observation in Fig. 1 of only very minor fractions of other phases.

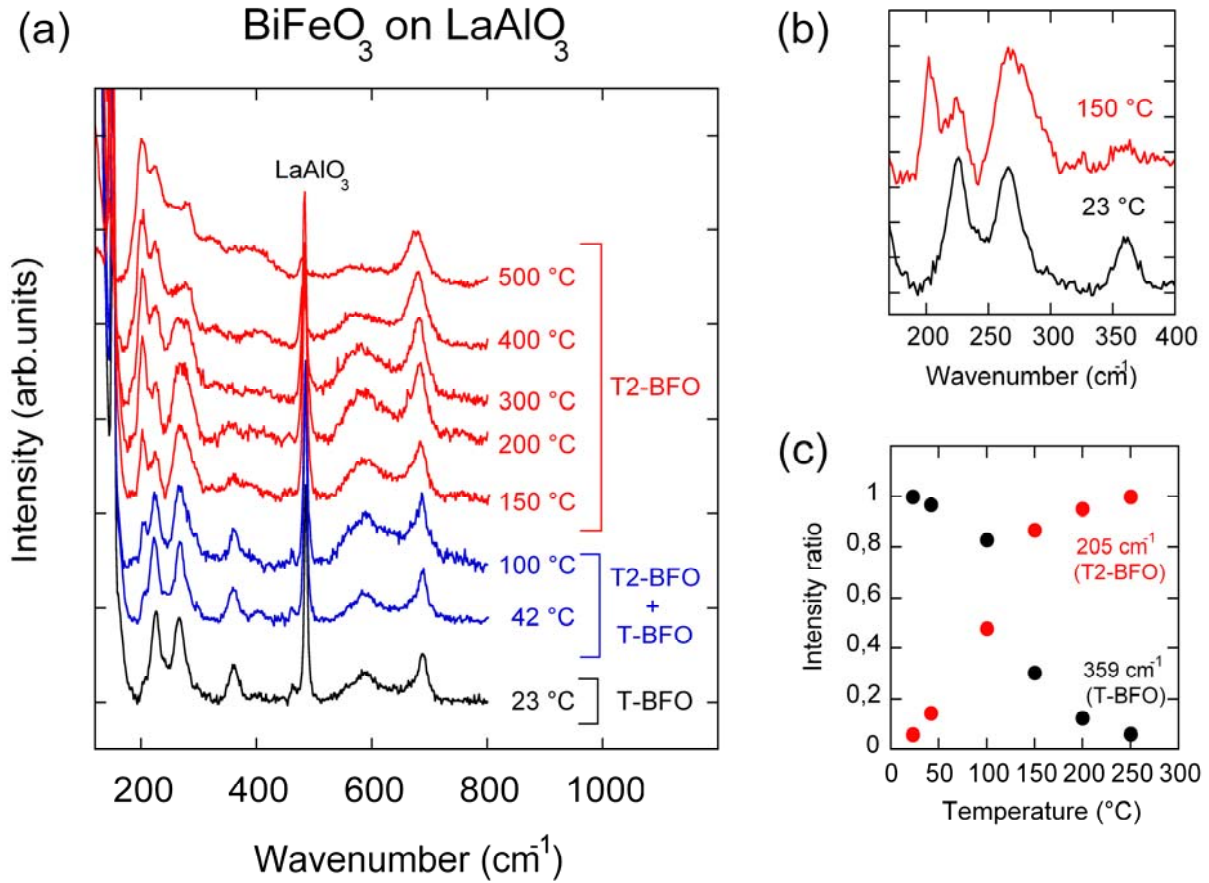

*Fig. 2 (colour on-line)*
(a) Selected temperature-dependent Raman spectra of a BFO thin film. The different colours of the spectra highlight the regime of the T-BFO phase (black), the coexistence regime between the T-BFO and T2-BFO phase (blue) and the signature of the new T2-BFO phase (red). Bands signed by LaAlO$_3$ correspond to Raman modes from the LAO substrate, where the low-wavenumber bands are cut for increased readability. (b) Comparison of the low-wavenumber region for a spectrum at 23 and 150 °C. (c) Evolution of the intensity ratio for the characteristic bands at 205 cm$^{-1}$ (T2-BFO) and 359 cm$^{-1}$ (T-BFO), the intensity ratio is normalized to one for the highest intensity in the shown temperature window.

We now discuss the effect of increasing temperature from 25 °C to 500 °C. The sharp substrate LAO bands follow the known continuous temperature evolution of its rhombohedral-to-cubic phase transition [28], i.e. the $A_{1g}$ mode at 120 cm$^{-1}$ softens to low wavenumber out of the spectral window and the modes at 150 and 484 cm$^{-1}$ decrease in intensity. The reduced intensity of the 150 and 484 cm$^{-1}$ LAO mode with increasing temperature also mirrors the decreasing transparency of BFO with increasing temperature [29]. Contrarily to the continuous changes of LAO, the temperature-dependent Raman spectra of the T-BFO film display three significant changes in the range from 42 to 150 °C: (*i*) the appearance of a new sharp band around 205 cm$^{-1}$, (*ii*) a shoulder with increasing intensity on the high-wavenumber wing of the 266 cm$^{-1}$ band, and (*iii*) a gradual disappearance of the initially strong band at 359 cm$^{-1}$. These important changes in the spectral signature are further illustrated by the panel (b) of Figure 2 which compares a low-wavenumber zoom for the spectra at 23 and 150 °C. Such significant changes provide experimental evidence for a structural change in T-BFO films as a function of temperature towards another phase. We note that the new spectral signature is different from the commonly observed spectrum for the *R*3*c* phase in single crystals or films [9, 23], so that a temperature-induced transition towards a relaxed *R*3*c* R-BFO film can be excluded. Fig. 2 shows that the T-BFO characteristic band around 685 cm$^{-1}$ is maintained at high-temperature; we thus hypothesise that the high-temperature phase is also characterised by a high *c*/*a* ratio. In the following, we will call this new phase T2-BFO. Let us now consider different points and possible scenarios for a better understanding of the transition from T-BFO to the T2-BFO phase.

The disappearance of the 359 cm$^{-1}$ band shows that the spectral signature above 150°C cannot be explained by a coexistence of the T-BFO phase with a new phase, but that we have to consider that the T-BFO phase entirely transits to the new T2-BFO phase. The spectra at 42 °C and 100 °C of Figure 2 are characterised by a coexistence of the bands at 359 cm$^{-1}$ (T-BFO signature) and at 205 cm$^{-1}$ (T2-BFO signature), as further illustrated by the intensity ratio shown in panel (c) of Fig. 2 . Such a phase coexistence is characteristic of a first-order phase transition.

The recent work by Diéguez *et al.* [8] suggests an intrinsic richness of structural phases in BFO by performing a systematic search for potentially stable phases by using first-principles methods. Diéguez *et al.* [8] have demonstrated that BFO can present an unusual large number of (meta)stable structures very close in energies. Notably, the calculations reveal that structures with a large *c*/*a* ratio present a particular intricate energy landscape with many possible phases such as *Pc*, *Cm*, *Pna*2$_1$ or *Cc* structures. All these structure have their energy minimum at a misfit strain of about -4.8%, implying that any of them can form a stable BFO film with epitaxial strain corresponding to a LAO substrate [8]. Because these phases are so close in energy, Diéguez *et al.* [8] have considered that the actual phase in thin films may depend on subtle experimental details. For our case of BFO films grown on (001) LAO substrates, the calculations suggest that the *Pc* and *Cm* phase are particularly close in energy. It is interesting to note that there exists no group-sub-group relationship between *Pc* and *Cm* so that a transition between these two phases is expected to be of first order, as observed through our Raman data.

It is also interesting to remind how the *R*3*c* bulk phase of BFO behaves under isotropic deformations (hydrostatic pressure). While it is known that the R3c structure is stable up to very high temperatures [4, 8, 29, 30], it has been shown that it is destabilized under hydrostatic pressure at a relatively modest pressure of 3 GPa [18, 31] and that new phases can easily be induced by non-hydrostatic conditions [32]. Although the biaxial strain in thin films cannot be directly compared to (pseudo-)hydrostatic conditions, such experiments add further support to the idea of a complex deformation-temperature phase diagram. Our temperature-dependent investigation of a highly strained film explores this complex phase diagram.

In summary, we have presented the synthesis, characterization, and temperature-dependent Raman scattering investigation of highly strained BFO thin films. The structural characterisation shows that the here investigated BFO thin films are largely dominated by the so-called supertetragonal T-BFO phase, with only minor traces of a complex mixture of secondary R- and $M_R$-phases. A temperature-dependent Raman scattering study of T-BFO films provides evidence for a first-order structural phase transition close to room temperature towards a new T2-BFO phase. We expect that the structural phase transitions lead to modified magnetic and ferroelectric properties and that the occurrence of a phase transition near room temperature allows enhanced properties and coupling in the model multiferroic $BiFeO_3$. During the submission of this manuscript we became aware of unpublished results [33, 34] that further encourage work in this direction.

*Acknowledgements*

P. Jadhav acknowledges an Erasmus Mundos Postdoctoral Fellowship within the External Cooperation Window program. Financial support by the Ministerio de Ciencia e Innovación of the Spanish Government [Projects MAT2008-06761-C03, MAT2011-29269-CO3 and NANOSELECT CSD2007-00041] and Generalitat de Catalunya (2009 SGR 00376) is acknowledged. Enlightening discussions with J. Iñiguez and O. Diéguez are also warmly acknowledged.